\documentclass[11pt,a4paper]{article}
\usepackage[marginratio=1:1,height=600pt,width=460pt,tmargin=100pt]{geometry}
\usepackage[utf8]{inputenc}
\usepackage[parfill]{parskip}    % Activate to begin paragraphs with an empty line rather than an indent
\usepackage{url}            % simple URL typesetting
\usepackage{booktabs}       % professional-quality tables
\usepackage{amsfonts}       % blackboard math symbols
\usepackage{nicefrac}       % compact symbols for 1/2, etc.
\usepackage{microtype}      % microtypography
\usepackage[dvipsnames]{xcolor}
\usepackage{graphicx}
\usepackage{amsmath}
\usepackage{amssymb}
\graphicspath{{./pics/}}
\usepackage{subcaption}
\usepackage{amsthm}
\usepackage{amscd}
\usepackage{mathrsfs}
\usepackage{bm}
\usepackage[linesnumbered,ruled,vlined]{algorithm2e}
\usepackage{multirow}
\usepackage{mathtools}
\usepackage{hyperref}
\usepackage{url}
\usepackage{enumitem}
\usepackage{authblk}
\usepackage{overpic}
\usepackage[normalem]{ulem}
\mathtoolsset{showonlyrefs}  % only show referred equation

\allowdisplaybreaks

\newcommand{\R}{\mathbb{R}}

\newcommand{\CL}{\mathcal{L}}
\newcommand{\bx}{\mathbf{x}}
\newcommand{\bp}{\mathbf{p}}
\newcommand{\bq}{\mathbf{q}}
\newcommand{\btheta}{\bm{\theta}}
\newcommand{\bxi}{\bm{\xi}}
\newcommand{\ct}{\mathcal{T}}
\newcommand{\cv}{\mathcal{V}}

\newtheorem*{thm*}{Theorem}

\newtheorem*{lemma*}{Lemma}

\usepackage{array}
\newcolumntype{C}[1]{>{\centering\let\newline\\\arraybackslash\hspace{0pt}}m{#1}}

\title{Data-Driven Discovery of Conservation Laws from Trajectories via Neural Deflation}
\author[1]{Shaoxuan Chen\footnote{Email: shaoxuanchen@umass.edu.}}
\author[1]{Panayotis G.~Kevrekidis}
\author[1]{Hong-Kun Zhang}
\author[2]{Wei Zhu}

\affil[1]{Department of Mathematics and
Statistics, University of Massachusetts
Amherst, Amherst, MA 01003-4515, USA}
\affil[2]{School of Mathematics, Georgia Institute of Technology, Atlanta, GA 30332, USA
}

\begin{document}

\maketitle

\begin{abstract}
In an earlier work by a subset of the present authors~\cite{neural_deflation}, the method of the
so-called neural deflation was introduced towards identifying
a complete set of functionally
independent conservation laws of a nonlinear dynamical system.
Here, we extend by a significant step this proposal. Instead of
using the explicit knowledge of the underlying equations of
motion, we develop the method directly from system trajectories.
This is crucial towards enhancing the practical implementation of
the method in scenarios where solely data reflecting discrete snapshots of
the system are available.
We showcase the results of the method and the number of associated
conservation laws obtained in a diverse range of examples
including 1D and 2D harmonic oscillators, the Toda lattice,
the Fermi-Pasta-Ulam-Tsingou lattice and the Calogero-Moser system.
\end{abstract}

\section{Introduction}

The discovery of conservation laws and the potential integrability of dynamical systems are crucial for understanding their evolution and provide valuable insights into their long-term behavior~\cite{goriely2001integrability}. Identifying conserved quantities is important in a wide range of domains, including optics~\cite{kivshar2003optical}, atomic physics~\cite{stringari}, materials science~\cite{dauxois}, fluid dynamics~\cite{infeld}, and plasma physics~\cite{plasmabook}, as it helps to characterize complex phenomena within these areas. Analyzing and uncovering these properties not only illuminates the underlying physical principles but also improves our capacity to model and predict the intricate behaviors of such systems, making this a key focus of ongoing scientific research.

Over the years, numerous theoretical and computational methods have been proposed for identifying the system's integrability~\cite{goriely2001integrability}. A partial subset of these
includes Lax pairs~\cite{ablowitz2}, the Painl{e}v{\'e} property~\cite{YAO2004723, conte}, Lyapunov exponents~\cite{benettin:1980a, benettin:1980b, sandra}, and Koopman operator theory~\cite{koopman1931hamiltonian, mezic2005spectral}. Recently, there has been a significant surge of interest in the data-driven discovery of conservation laws and the assessment of system integrability using machine learning; see for some prominent examples, e.g., the works of~\cite{kaiser2018discovering, siam, teg1, arora2023model, lu2023discovering, liu2022machine, liu2024interpretable, neural_deflation, bondesan2019learning, ha2021discovering} and the references therein. Despite extensive efforts and advancements in this field, most of these methods still face nontrivial limitations. Many are, in practice, capable of uncovering only a single or a few conservation laws, even for integrable systems, especially so for systems
with many degrees of freedom~\cite{siam, ha2021discovering}. Methods that purport to discover a comprehensive set of independent conservation laws often either fail in practice or require substantial parameter tuning and human interpretation, reducing their robustness and reliability~\cite{kaiser2018discovering, teg1, lu2023discovering, liu2022machine, liu2024interpretable}. Moreover, several approaches necessitate explicit knowledge of the ordinary differential equations (ODEs) governing the dynamics or the precise number of independent conservation laws within the system, thereby limiting their applicability in practical scenarios where only system trajectories may be available in the form of dynamical evolution obtained (or, e.g., experimentally acquired) ``data''~\cite{liu2022machine, liu2024interpretable, neural_deflation}.

Accordingly, in this paper, we propose a data-driven method for discovering a \textit{complete} set of \textit{functionally independent} conservation laws from system trajectories. A key component of our approach is the, so-called, neural deflation technique~\cite{neural_deflation}, previously introduced by some of the authors. Unlike the original method, however, our approach does not require explicit knowledge of the underlying ODEs and relies solely on observed system trajectories. This significantly enhances its practicality in real-world scenarios where the exact dynamics are unknown, and only discrete snapshots of the system are available. We demonstrate the effectiveness of our method on widely benchmarked systems, such as the 1D and 2D harmonic oscillators, as well as on integrable and non-integrable differential-difference equation 
models, including the Toda lattice~\cite{toda1989theory}, the Fermi-Pasta-Ulam-Tsingou (FPUT) system~\cite{gallavotti2007fermi}, the Calogero-Moser model~\cite{10.1063/1.1665604, MOSER1975197}, and the discrete sine-Gordon equation~\cite{dauxois}.  Importantly, we highlight both positive and negative results, illustrating the pitfalls that may arise when working with system trajectories instead of ground-truth ODEs, an aspect that has been, in our view,  both underexplored and underreported in previous research.

The rest of the paper is structured as follows. Section~\ref{sec:related} discusses related works and their limitations, setting the stage
for the present contribution. In Section~\ref{sec:background}, we provide a brief overview of the theoretical toolbox background on conservation laws, the notion of (Liouville) integrability, Hamiltonian neural networks, and the neural deflation method. Section~\ref{sec:method} details our data-driven approach for learning a complete set of conservation laws from system trajectories. In Section~\ref{sec:experiments}, we present a series of numerical experiments on various systems, demonstrating the effectiveness of our method and highlighting its potential pitfalls. Finally, we conclude the paper in Section~\ref{sec:conclusion} and discuss directions for future work.

\section{Related Works}
\label{sec:related}
In recent years, there has been a growing focus on using data-driven methods to identify conservation laws and assess system integrability through various machine learning techniques. These approaches fall into two categories: those utilizing explicit knowledge of the underlying ODEs and those relying solely on observed system trajectories.

In the first category, where the underlying ODEs are known, the most recent state-of-the-art methods are described in \cite{liu2022machine, liu2024interpretable, neural_deflation}. \cite{liu2022machine} initiated this approach by using a regularized loss function to train a collection of neural networks that parameterize conservation laws, aiming to promote---but not necessarily guarantee---their functional independence. \cite{liu2024interpretable} further developed this method by incorporating sparse regression to improve the interpretability of the learned conserved quantities, utilizing a known set of basis functions. Only after a pre-determined set of conservation laws is learned, which are, however, often \textit{neither complete nor functionally independent}, a maximal subset of independent conservation laws is then selected. Consequently, these methods generally cannot ensure that a complete set of independent conservation laws is identified. Indeed, in practical applications to integrable systems, these approaches typically uncover only a few conserved quantities \cite{liu2022machine}, which are argued to be physically relevant. The neural deflation method \cite{neural_deflation} adopts a different approach by iteratively constructing a sequence of deflated loss functions, ensuring that each newly identified conserved quantity is independent of those found earlier. This method has been shown to successfully discover a complete set of functionally independent conservation laws in various systems. Nonetheless, this method still depends on explicit knowledge of the underlying ODE, and its effectiveness in accurately determining the exact number of conserved quantities based solely on system trajectories remains uncertain.

In the second category, which assumes only the availability of system trajectories, the methods tend to be less robust and reliable, often requiring significant parameter tuning, human interpretation, or additional assumptions. The Siamese neural networks proposed by Wetzel et al.~\cite{PhysRevResearch.2.033499} can only learn a single conservation law. While the method proposed by Ha and Jeong~\cite{ha2021discovering} can theoretically learn multiple conserved quantities, it requires ``grouped data" sampling from level sets of previously identified conserved quantities, which significantly limits its practical applicability. The model-agnostic learning technique by Arora et al.~\cite{arora2023model} necessitates explicit knowledge of the exact number of independent conservation laws in the system. For manifold learning approaches~\cite{teg1, lu2023discovering}, there is typically no guarantee that the correct dimension of the isosurface of the conserved quantities---and therefore the correct number of conservation laws---will be learned and identified. The aim of the present work is to combine the
methodological advatanges of~\cite{neural_deflation} with an implementation
that can work directly with data.

\section{Background and Motivation}
\label{sec:background}

This section provides a brief overview of the theoretical notions
of interest regarding Hamiltonian systems, their conservation laws, 
and the method of neural deflation.
%Hamiltonian systems, conservation laws, (Liouville) integrability, and the neural deflation method for data-driven learning of conservation laws and assessing system integrability.

\subsection{Hamiltonian Systems and Conservation Laws}
Consider a \textit{Hamiltonian system} of $d$ degrees of freedom,
\begin{equation}
     \frac{d\mathbf{x}}{dt} = \mathbf{f}(\mathbf{x}), \quad \mathbf{f}(\mathbf{x}) = J(\mathbf{x})\nabla H(\mathbf{x}), \quad \forall \mathbf{x}\in D \subset \mathbb{R}^{2d},
     \label{eq:d-hamiltonian}
  \end{equation}
where $H: D \to \mathbb{R}$ is the Hamiltonian function, and $J(\mathbf{x}) \in \mathbb{R}^{2d \times  2d}$ is an antisymmetrix matrix. Let $F$ and $G$ be two smooth functions over the phase space $D$, their \textit{Poisson bracket}, $\{F, G\}:D\to\R$, is given by
\begin{equation}
     \{F, G \}(\mathbf{x}) \coloneqq \nabla F(\mathbf{x})^T J(\mathbf{x}) \nabla G(\mathbf{x}), \quad \forall \mathbf{x}\in D.
     \label{eq: poisson bracket}
\end{equation}
In particular, under the canonical coordinates $\mathbf{x} = (\mathbf{q}, \mathbf{p})$, where $\mathbf{q},\mathbf{p}\in\R^d$ are the generalized positions and conjugate momenta, we have
\begin{align}
    J(\mathbf{x})\equiv
    \begin{bmatrix}
        0 & I_d\\
        -I_d & 0
    \end{bmatrix}, \quad
    \{F, G \} = \nabla_{\mathbf{q}}F\cdot \nabla_{\mathbf{p}}G - \nabla_{\mathbf{p}}F\cdot \nabla_{\mathbf{q}}G,
\end{align}
and the familiar Hamilton's equations are of the form:
\begin{align}
\label{eq:d-hamiltonian-canonical}
    \left\{
    \begin{aligned}
        &\frac{\mathrm{d}\bq}{\mathrm{d}t} = \nabla_{\bp} H,\\
        &\frac{\mathrm{d}\bp}{\mathrm{d}t} = -\nabla_{\bq} H.
    \end{aligned}\right.
\end{align}

A $C^1$ smooth function $I: D \to \mathbb{R}$ is called a \textit{conservation law} of system~\eqref{eq:d-hamiltonian} if it remains constant over system trajectories. That is, for any $\mathbf{x}(t)$ that is a solution of Eq.~\eqref{eq:d-hamiltonian}, we have
\begin{align}
    I(\mathbf{x}(t))\equiv I(\mathbf{x}(0)), \quad \forall t\geq 0.
\end{align}
It is easy to verify that $I$ is a conservation law of system~\eqref{eq:d-hamiltonian} if and only if its Poisson bracket with the Hamiltonian $H$ vanishes over $D$, i.e.,
\begin{equation}
     \{I, H \}(\mathbf{x}) = \nabla I(\mathbf{x}) \cdot \mathbf{f} (\mathbf{x}) = 0, \quad \forall \mathbf{x} \in D
     \label{eq: conservation_law}
\end{equation}

Let $\{I_k:D\to\R \}_{k=1}^K$ be a collection of $K$ conservation laws of system~\eqref{eq:d-hamiltonian}, we say they are \textit{functionally independent} if their gradients $\{ \nabla I_k(\mathbf{x}) \}_{k=1}^{K} $ are linearly
independent vectors in $\mathbb{R}^{2d}$ for almost every $\mathbf{x}\in D$. Intuitively, this means there is no non-trivial relationship among these conserved quantities that allows one to be expressed as a (nonlinear) combination of the others. Moreover, these conservation laws are said to be \textit{in involution}, or \textit{Poisson commuting},  if their pairwise Poisson brackets vanish, i.e., $\ \{ I_j,I_k \} = 0, \forall j \ne k$. For a Hamiltonian system with $d$ degrees of freedom, there can be at most $d$ functionally independent conservation laws in involution. When a system admits $d$ such conserved quantities, it is said to be completely integrable in the Liouville sense \cite{arnol2013mathematical}. 
%It is worth mentioning in 
%passing that there are other notions of integrability that we will not touch
%upon here, such as the
%Frobenius sense discussed, e.g., in~\cite{liu2022machine}.

\subsection{Neural Deflation Method}
\label{sec:neural_deflation}
The neural deflation method~\cite{neural_deflation} provides a systematic, data-driven approach for identifying a maximal set of functionally independent, Poisson-commuting conservation laws when the underlying dynamics, i.e., \(\mathbf{f}(\mathbf{x})\) in Eq.~\eqref{eq:d-hamiltonian}, is explicitly known. The core idea is to \textit{iteratively} learn each conservation law using a \textit{deflated} loss function, ensuring that each newly identified conserved quantity is both in involution and functionally independent from the previously discovered ones. This approach is inspired by the concept of ``deflation", which is used to find multiple steady states of partial differential equations by adjusting the loss function to discourage solutions that are similar to those already identified~\cite{patrick}.

The process begins by randomly sampling a training set \(\mathcal{T}\) and a validation set \(\mathcal{V}\) from \(D \subset \mathbb{R}^{2d}\). Each conserved quantity \(I_k(\mathbf{x})\) is parameterized using a neural network \(I_k(\mathbf{x}; \bm{\theta}_k)\), where \(\bm{\theta}_k\) are the trainable parameters. To learn the first conservation law \(I_1(\mathbf{x}; \bm{\theta}_1)\), the following loss function is minimized, based on the condition~\eqref{eq: conservation_law}:
\begin{align}
\label{eq:l1}
    \mathcal{L}_1(\bm{\theta}_1; \mathcal{T}) \coloneqq \frac{1}{|\mathcal{T}|} \sum_{\mathbf{x}\in \mathcal{T}} \left| \widehat{\mathbf{f}}(\mathbf{x}) \cdot \widehat{\nabla I_1 }(\mathbf{x};\bm{\theta}_1) \right|^{2},    
\end{align}
where \(\widehat{\mathbf{f}}(\mathbf{x})\) and \(\widehat{\nabla I_1 }(\mathbf{x};\bm{\theta}_1)\) are the \(l^2\)-normalized vectors {of \(\mathbf{f}(\mathbf{x})\) and \(\nabla I_1 (\mathbf{x};\bm{\theta}_1)\), respectively, evaluated on the training set $\mathcal{T}$.} This normalization makes the loss function scale-invariant and prevents learning the trivial conservation law \(I_1(\mathbf{x}; \bm{\theta}_1) \equiv C\).

The process continues by iteratively learning additional conservation laws inductively as follows. Assuming \(K-1\) conserved quantities \(\{I_k(\mathbf{x}; \bm{\theta}^*_k)\}_{k=1}^{K-1}\) have already been found, the \(K\)-th conservation law \(I_K(\mathbf{x}; \bm{\theta}_K)\) is trained using the following deflated loss function \(\mathcal{L}_K(\bm{\theta}_K; \mathcal{T})\), while keeping the parameters of the previous networks \(\{\bm{\theta}^*_k\}_{k=1}^{K-1}\) fixed:
\begin{align}
\label{eq:lK}
\mathcal{L}_K(\bm{\theta}_K; \mathcal{T}) \coloneqq \frac{1}{|\mathcal{T}|}\sum_{\mathbf{x}\in\mathcal{T}}\frac{\displaystyle
     \overbrace{\ell_{\text{conserv}}[\bm{\theta}_K; \mathbf{x}]}^{\text{conservation loss}} + \overbrace{\sum^{K-1}_{k=1} \ell_{\text{inv}}[\bm{\theta}^{*}_k, \bm{\theta}_K; \mathbf{x}]
     }^{\text{involution loss}}}{\displaystyle \underbrace{K \left| \ell_{\text{ind}}[\bm{\theta}_K|\bm{\theta}^{*}_1, \dots, \bm{\theta}^{*}_{K-1}; \mathbf{x}] \right|^{\alpha}}_{\text{independent loss}}},
\end{align}
where
\begin{itemize}
    \item The \textit{conservation loss}, $\ell_{\text{conserv}}[\bm{\theta}_K; \mathbf{x}]$, ensures that \(I_K(\mathbf{x}; \bm{\theta}_K)\) is indeed a conserved quantity, and is defined similarly to  Eq.~\eqref{eq:l1}:
    \begin{align}
        \ell_{\text{conserv}}[\bm{\theta}_K; \mathbf{x}] \coloneqq \frac{1}{|\mathcal{T}|} \sum_{\mathbf{x}\in \mathcal{T}} \left| \widehat{\mathbf{f}}(\mathbf{x}) \cdot \widehat{\nabla I_K }(\mathbf{x};\bm{\theta}_K) \right|^{2}.
    \end{align}
    \item  The \textit{involution loss}:
    \begin{align}
        \ell_{\text{inv}}[\bm{\theta}^{*}_k, \bm{\theta}_K; \mathbf{x}] \coloneqq \left| \{ I_{k} (\cdot ; \bm{\theta}^{*}_k),  I_{K} (\cdot ; \bm{\theta}_K) \}(\mathbf{x}) \right|^{2}
    \end{align}
    stipulates that \(I_K(\cdot; \bm{\theta}_K)\) is in involution with all previously identified conservation laws \(\{I_k(\mathbf{x}; \bm{\theta}^{*}_k) \}_{k=1}^{K-1}\).
    \item The \textit{independent loss} in the deflated denominator is defined as:
    \begin{align}
    \label{eq:ind_loss}
        \ell_{\text{ind}}[\bm{\theta}^{*}_k| \bm{\theta}_1, \cdots,\bm{\theta}_{K-1};\mathbf{x}] \coloneqq \left\|\text{Proj}_{\text{span}\{\widehat{\nabla I_k}(\mathbf{x}; \bm{\theta}^{*}_k)  \}_{k\in[K-1]}^{\perp}} \widehat{\nabla I_K}(\mathbf{x}; \bm{\theta}_K)\right\|^{2},
    \end{align}
    where \(\text{Proj}_{\text{span}\{\widehat{\nabla I_k}(\mathbf{x}; \bm{\theta}^{*}_k) \}_{k\in[K-1]}^{\perp}} \widehat{\nabla I_K}(\mathbf{x}; \bm{\theta}_K)\) represents the projection of \(\widehat{\nabla I_K}(\mathbf{x}; \bm{\theta}_K)\) onto the orthogonal complement of the subspace spanned by \(\{ \widehat{\nabla I_k}(\mathbf{x}; \bm{\theta}^{*}_k)\}_{k\in[K-1]}\) in $\R^{2d}$. This term introduces a singularity that penalizes any lack of independence between \(\widehat{\nabla I_K}(\mathbf{x}; \bm{\theta}_K)\) and the previously learned \(\{ \widehat{\nabla I_k}(\mathbf{x}; \bm{\theta}^{*}_k)\}_{k\in[K-1]}\). The deflation parameter \(\alpha > 0\) is a hyperparameter that controls the strength of this functional independence constraint between \(I_K(\cdot; \bm{\theta}_K)\) and \(\{I_k(\mathbf{x}; \bm{\theta}^{*}_k) \}_{k=1}^{K-1}\).
\end{itemize}
One can easily show that this method is \textit{consistent} in the infinite-sample limit in the following sense: if the previously obtained $\{I_k(\cdot;\btheta_k^*)\}_{k=1}^{K-1}$ accurately parameterize a ground-truth set of independent conservation laws in involution, and if the empirical averages over the training set $\ct$ are replaced by expectations with respect to an absolutely continuous probability measure, then the loss $I_K(\cdot; \btheta_K^*)$ is zero \textit{if and only if} $\{I_k(\cdot;\btheta_k^*)\}_{k=1}^{K}$ forms a new set of $K$ independent Poisson-commuting conservation laws. For more details, see the proof provided in~\cite{neural_deflation}.

The process is repeated until a significant increase in the loss function $\mathcal{L}_K(\btheta_K^*; \cv)$ is observed on the validation set $\cv$. At that point, $\{I_k(\cdot; \btheta_k^*)\}_{k=1}^{K-1}$ is declared  a maximal set of $d_0=K-1$ independent conservation laws in involution for the system.

Given the expressiveness of neural networks and the availability of the ground-truth Hamiltonian $H(\bx)$, it is feasible to sample an extensive training set $\mathcal{T}$ from $D$ to construct the consistent loss~\eqref{eq:lK} for training any conserved quantities. However, the effectiveness of this neural deflation method becomes uncertain when direct access to $H(\bx)$ is not available, and only a limited set of observed or simulated trajectories can be used. This challenge is the central focus of this work.

\section{Method}
\label{sec:method}

A natural strategy to address this problem is to infer the Hamiltonian of the system from its observed trajectories. Since the loss functions~\eqref{eq:l1} and~\eqref{eq:lK} do not require a closed-form expression for the Hamiltonian but only the ability to sample sufficient values of the vector field $\mathbf{f}(\bx) = J(\bx)\nabla H(\bx)$ for $\bx \in \mathcal{T} \subset D$, a numerical approximation of the ground-truth Hamiltonian suffices. This approach naturally raises two questions:

\begin{itemize}
    \item How can we approximate the Hamiltonian using only observed or simulated system trajectories, and what is the expected accuracy of this approximation?
    \item If we use a numerically approximated Hamiltonian in place of the exact Hamiltonian in the neural deflation method, will the method still function effectively?
\end{itemize}

\subsection{Hamiltonian Neural Network}
There is a substantial body of research on data-driven learning of Hamiltonian systems~\cite{toth2019hamiltonian, bertalan2019learning, HNN, PhysRevResearch.3.023156, gupta2023hamiltonian, mattheakis2022hamiltonian, bienias2021meta, david2023symplectic, chen2019symplectic, dipietro2020sparse}. Among these, the Hamiltonian Neural Network (HNN)~\cite{HNN} stands out as one of the pioneering approaches. Several improvements to HNNs have been proposed, including the use of more accurate integrators~\cite{david2023symplectic}, enhanced modeling of continuous-time trajectories~\cite{haitsiukevich2022learning}, and adaptation to various bifurcation parameters~\cite{PhysRevResearch.3.023156}. In this paper, we will adopt the original HNN~\cite{HNN} to derive a neural network approximation of the Hamiltonian from observed system trajectories. This choice is made both for its simplicity and to facilitate the evaluation of whether a numerically approximated Hamiltonian (with inherent uncertainties) can effectively support the downstream neural deflation learning of the remaining conservation laws.

For simplicity, let us assume that the Hamiltonian system is observed using the canonical coordinates \(\bx = (\bq, \bp) \in \mathbb{R}^{2d}\). The dataset consists of a collection of \(I\) trajectories. Note that here, we slightly abuse the notation $I$, which was previously used 
with a subscript to denote/index the conservation laws, to also represent the number of trajectories in the data. Each trajectory is sampled at \(J\) time steps with a uniform time interval \(\Delta t\).
\begin{align}
\label{eq:data_D}
    \mathcal{D} = \left\{\bx^{(i,j)} = \left(\bq^{(i,j)}, \bp^{(i,j)}\right) : 1\le i \le I, ~1\le j\le J\right\},
\end{align}
where $\textbf{q}^{(i,j)} = (q_1^{(i,j)}, \cdots,q_d^{(i,j)})\in\R^d$ represents the generalized position coordinates, and $\textbf{p}^{(i,j)} = (p_1^{(i,j)}, \cdots,p_d^{(i,j)})$ denotes the conjugate momenta. Here, $i\in\{1, \cdots, I\}$ indexes the sampled trajectories, and $j\in \{1, \cdots, J\}$ represents the time index within each trajectory.

Assuming a small time interval $\Delta t$, one can estimate the time derivative of the observed trajectories using finite differences, yielding the time-derivative dataset:
\begin{align}
\label{eq:derivate_data}
    \dot{\mathcal{D}} = \left\{\dot{\bx}^{(i,j)} = \left(\dot{\bq}^{(i,j)}, \dot{\bp}^{(i,j)}\right) : 1\le i \le I, ~1\le j\le J\right\}.
\end{align}
{For high-accuracy time-derivative approximations, we typically use 8th-order central finite differences. However, experiments show that lower-order approximations, though less precise, generally yield qualitatively similar final outcomes for our method.}

We can then parameterize the unknown Hamiltonian $H(\bq, \bp)$ using an HNN, denoted as $H_{\bm{\xi}}(\bq, \bp)$, where $\bxi$ represents the network parameters. The HNN is trained using the following loss function based on the Hamiltonian dynamics~\eqref{eq:d-hamiltonian-canonical}:
\begin{align}
\label{eq:l2_hnn}
        \mathcal{L}_{\text{HNN}}(\bxi; \mathcal{D}, \dot{\mathcal{D}})= \frac{1}{IJ}\sum_{i,j}\Big\| \nabla_{\bp} H_{\bxi}\left(\bq^{(i,j)}, \bp^{(i,j)}\right) - \dot{\bq}^{(i,j)} \Big\|^2 + \Big\|  \nabla_{\bq}H_{\bxi}\left(\bq^{(i,j)}, \bp^{(i,j)}\right) + \dot{\bp}^{(i,j)}\Big\|^2,
\end{align}
One solves the above problem through optimization, where the gradients $(\nabla_{\bq}H_{\bxi}, \nabla_{\bp}H_{\bxi})$ are obtained via automatic differentiation. The resulting minimizer $\bxi^*$ provides a parameterized approximation of the Hamiltonian, $H_{\bxi^*}(\bq, \bp)$.

\subsection{Neural Deflation with HNN}
\begin{figure}[h!]
  \centering
  \includegraphics[width=1.0\linewidth]{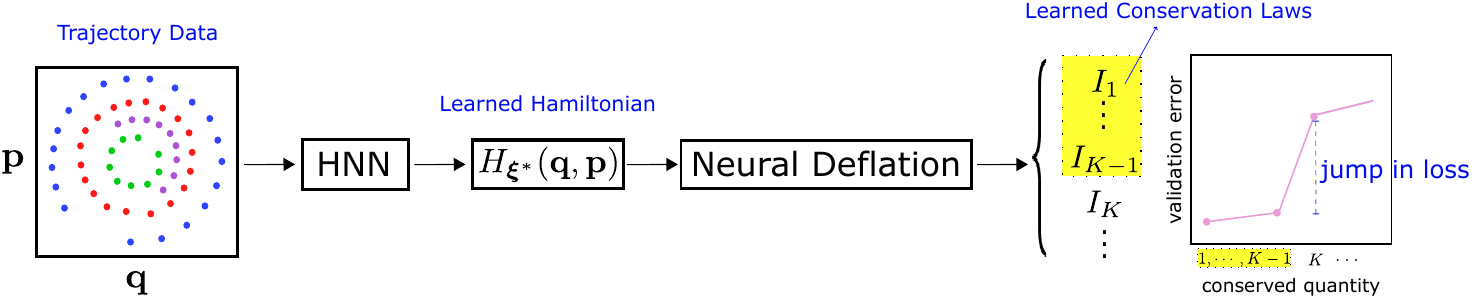}
   \caption{{Schematic representation of our method. Trajectory data (with different colors representing distinct trajectories) are first input into the Hamiltonian Neural Network (HNN) to learn the Hamiltonian \(H_{\bm{\xi}^*}(\bq, \bp)\), which is then incorporated into the neural deflation method. A jump in the validation loss at \(I_K\) signals the successful learning of \(K-1\) Poisson-commuting, functionally independent conservation laws.}}
   \label{fig:pipleline}
\end{figure}

Once the neural network approximation \(H_{\bm{\xi}^*}(\bq, \bp)\) of the Hamiltonian is obtained, it can be incorporated into the neural deflation method described in Section~\ref{sec:neural_deflation}. This involves replacing the unknown Hamiltonian \(H(\bx)\) and the vector field \(\mathbf{f}(\bx)\) with their numerically approximated counterparts, \(H_{\bm{\xi}^*}(\bx)\) and \(\mathbf{f}_{\bm{\xi}^*}(\bx)\), respectively. Our method is summarized in Algorithm~\ref{alg:deflation}, {with a schematic representation presented in Figure~\ref{fig:pipleline}}.

\begin{algorithm}[h]
  \SetAlgoLined
  \KwIn{$ \mathcal{D} = \left\{\bx^{(i,j)} : 1\le i \le I, ~1\le j\le J\right\}\subset D\subset \R^{2d}$, a collection of \(I\) trajectories of a Hamiltonian system, each sampled at \(J\) time steps with a uniform time interval \(\Delta t\).}
  \KwOut{Neural network approximation $H_{\bxi}(\bx)$ of the Hamiltonian. A maximal set $\{I_k(\cdot; \btheta_k^*)\}_{k=1}^{d_0}$ of functionally independent, Poisson-commuting conservation laws. }
  Estimate the time-derivatives $\dot{\mathcal{D}}=\left\{\dot{\bx}^{(i,j)} : 1\le i \le I, ~1\le j\le J\right\}$~\eqref{eq:derivate_data} using finite differences\;
  $\bxi^* \leftarrow \arg\min_{\bxi}\CL_{\text{HNN}}(\bxi; \mathcal{D}, \dot{\mathcal{D}})$ given by Eq.~\eqref{eq:l2_hnn}, and use the trained HNN, $H_{\bxi^*}(\bx)$, as a surrogate Hamiltonian for all subsequent steps\;
  Randomly sample a training set $\ct$ and a validation set $\cv$  from the phase space $D\subset \R^{2d}$\;
  $\btheta_1^* \leftarrow \arg\min_{\btheta_1}\CL_1(\btheta_1; \ct)$ given by Eq.~\eqref{eq:l1}\;
  $\CL_1^\text{val} \leftarrow \CL_1(\btheta_1^*; \cv)$\;
  $K\leftarrow 1$\;
  \Repeat{$\CL_K^{\text{\normalfont val}}/\CL_1^{\text{\normalfont val}} > \text{\normalfont tol}$}{
    $K\leftarrow K+1$\;
    $\btheta_K^* \leftarrow \arg\min_{\btheta_K}\CL_K(\btheta_K; \ct)$ given by \eqref{eq:lK}\;
    $\CL_K^\text{val} \leftarrow \CL_K(\btheta_K^*; \cv)$\;
  }
  $d_0\leftarrow K-1$\;
  \caption{Neural deflation with HNN}
  \label{alg:deflation}
\end{algorithm}

In practice, \(H_{\bm{\xi}^*}(\bq, \bp)\), the trained HNN, provides only a numerical approximation of the true Hamiltonian \(H(\bq, \bp)\), which inevitably introduces inaccuracies, especially in regions of the domain \(D\) that are not covered by the sampled trajectories. This work seeks to carefully evaluate these inaccuracies and their effect on the neural deflation learning of the remaining conservation laws.

\section{Numerical Experiments}
\label{sec:experiments}

In this section, we present the results of our algorithm in learning independent conservation laws from the simulated trajectories of various examples, including the 1D and 2D harmonic oscillators, the discrete sine-Gordon system, the Calogero-Moser system, the integrable Toda lattice, and the nonintegrable Fermi-Pasta-Ulam-Tsingou system.

\begin{itemize}
    \item \textbf{1D and 2D harmonic oscillators}. We assume the \textit{1D harmonic oscillator} to be describing the motion of a mass \(m\) attached to a spring with a spring constant \(k\). 
    %It is a single-degree-of-freedom Hamiltonian system represented by the 
    The relevant canonical coordinates are \(\bx = (q, p) \in \mathbb{R}^2\), where \(q\) is the position and \(p\) is the momentum. The equation of motion is given by:
    \begin{align}
         \dot{q} = \frac{p}{m}, \quad \dot{p} = -kq.
        \label{eq:1d_harmonic_oscillator}    
    \end{align}
    The Hamiltonian for the 1D harmonic oscillator consisting of the sum
    of kinetic and potential energies reads:
    %representing the total energy of the system, consists of the kinetic energy of the mass and the potential energy stored in the spring:
    \begin{align}
    \label{eq:hamiltonian_1d_osci}
        H = \frac{p^2}{2m} + \frac{1}{2}kq^2.
    \end{align}
    The \textit{2D harmonic oscillator} generalizes the motion to two dimensions, described by the coordinates \(\bx = (\bq, \bp)\), where \(\bq = (q_1, q_2)\) and \(\bp = (p_1, p_2)\). The corresponding Hamiltonian for the 2D system is:
    \begin{align}
    \label{eq:hamiltonian_2d_osci}
        H = \frac{1}{2}k_1q_1^2 + \frac{p_1^2}{2m_1} + \frac{1}{2}k_2q_2^2 + \frac{p_2^2}{2m_2}
    \end{align}
    Both the 1D and 2D harmonic oscillators are simple, integrable systems, making them natural examples for a preliminary assessment of the accuracy of our method. We will use these systems to demonstrate how effectively the HNN can learn the true Hamiltonian from observed trajectories and evaluate its impact on the subsequent neural deflation learning of conservation laws.
\end{itemize}

We also consider discrete Hamiltonian lattice systems consisting of \(N\) identical nodes, which can potentially be large in number, with dynamic variables \((\bq, \bp) = (q_1, \ldots, q_N, p_1, \ldots, p_N) \in \mathbb{R}^{2N}\). 
While numerous Hamiltonian systems bear a different symplectic structure
and a different form of the Hamiltonian, the present form constitutes a
large enough class of problems of relevance to applications (as is reflected
in the citations below) that merits, in our view, the present investigation.
Generalizations to other types of Hamiltonian systems are worthwhile
to consider in their own right and are deferred to future work.
The systems we examine all have a Hamiltonian function of the following form:
\begin{align}
    H(\textbf{q, p}) = T(\textbf{p}) + V(\textbf{q}),
    \label{eq:discrete_hamiltonian_lattice}
\end{align}
where \(T(\textbf{p}) = \sum_{n=1}^{N} \frac{p_n^2}{2}\) represents the kinetic energy, and \(V(\textbf{q})\) denotes the potential energy. We assume periodic boundary conditions, so \(q_{N+1} = q_1\). By selecting different forms of the potential \(V(\textbf{q})\), these systems can be either integrable or non-integrable.

\begin{itemize}
    \item \textbf{Discrete sine-Gordon system.} The discrete sine-Gordon system consists of a one-dimensional lattice of particles connected by nonlinear springs \cite{braun,Sine_Gordon_Model_Applications}. The Hamiltonian for this system is defined as in \eqref{eq:discrete_hamiltonian_lattice}, with a nonlinear potential:
    \begin{equation}
         V(\textbf{q}) = \sum_{n=1}^{N} \left[ \frac{C}{2}(q_{n+1} - q_n)^2 + 1 - \cos q_n \right],
         \label{eq:discrete_sine_gordon}
    \end{equation}
    where \(C\) is a small nonlinear coupling parameter. In this model, the
    above mentioned energy is the only conserved quantity.
    \item \textbf{Calogero-Moser system.}  The Calogero-Moser system is a classical mechanics model describing particles that interact via a pairwise potential inversely proportional to the square of their relative distance \cite{MOSER1975197,10.1063/1.1665604}:
    \begin{equation} V(\bq) = \sum_{l<k} \frac{1}{(q_k - q_l)^2} \label{eq:Calogero} \end{equation}
    This system is fully integrable, meaning it has $N$ independent conserved quantities that are in involution. These conserved quantities are directly associated with the underlying symmetries of the Hamiltonian~\cite{MOSER1975197}.
    \item \textbf{Toda lattice}. The Toda lattice \cite{toda1989theory} models the dynamics of a nonlinear one-dimensional crystal. It describes the motion of a chain of particles interacting through exponential forces that depend on the distance between nearest neighbors. The potential energy of the system is given by:
    \begin{equation}
         V(\textbf{q}) = \sum_{n=1}^{N} [e^{q_n-q_{n-1}}+ (q_{n+1}-q_n)-1]
         \label{eq:toda_lattice}
      \end{equation}
    The Toda lattice is also fully integrable, possessing $N$ conservation quantities. These conservation laws are explicitly derived in the classic study by \cite{PhysRevB.9.1921}.
    \item \textbf{Fermi-Pasta-Ulam-Tsingou system}. Finally, the Fermi-Pasta-Ulam-Tsingou (FPUT) system is closely associated with the Toda lattice \cite{osti_4376203,gallavotti2007fermi}, resembling a leading order
    Taylor-expansion thereof. Its nonlinear potential is:
    \begin{equation}
         V(\textbf{q}) = \sum_{n=1}^{N}  (q_n-q_{n+1})^2/2-(q_{n+1}-q_n)^{3}/6
         \label{eq:FPUT}
    \end{equation}
    However, unlike the Toda lattice~\eqref{eq:toda_lattice}, the FPUT system is non-integrable. In addition to conserving the Hamiltonian, the
    relevant model only conserves the momentum, $P =  \sum_{n=1}^{N} p_n$.
\end{itemize}

\subsection{Data Generation and Hyperparameters}
\label{sec: data_generation}
The dataset \(\mathcal{D} = \left\{\bx^{(i,j)} = \left(\bq^{(i,j)}, \bp^{(i,j)}\right) : 1 \le i \le I, ~1 \le j \le J\right\}\)~\eqref{eq:data_D} is generated by numerically integrating the systems using the 8th-order Runge-Kutta method~\cite{book_dop}. For simplicity, all initial conditions are uniformly sampled from a Euclidean box \([-r, r]^{2d}\). Table~\ref{tab:data generation for systems} lists the parameters used for simulating each system, including:
\begin{itemize}
    \item \(I\): the number of trajectories.
    \item \(J\): the number of time steps in each trajectory.
    \item \(T\): the terminal time of the simulation.
    \item \(r\): the size of the sampling box.
\end{itemize}

To learn the Hamiltonian of the system from $\mathcal{D}$, we first estimate the time-derivative data $\dot{\mathcal{D}} = \left\{\dot{\bx}^{(i,j)} = \left(\dot{\bq}^{(i,j)}, \dot{\bp}^{(i,j)}\right) : 1\le i \le I, ~1\le j\le J\right\}$ using the 8th-order central differences. A seven-layer fully-connected neural network, with 1000 hidden units per layer and $\tanh$ activation function, is employed to parameterize the HNN. The network is trained using the Adam optimizer~\cite{adam} on random batches of size 32. The initial learning rate is set to $10^{-5}$ and training continues for 40,000 steps, after which the learning rate is reduced to $10^{-9}$ for an additional 20,000 steps to fine-tune the HNN and ensure convergence.  A detailed summary of the hyperparameters is provided in Table~\ref{tab:HNNs_setting}. {Our method is relatively robust to changes in hyperparameters, so slight deviations from the chosen set of values do not lead to qualitatively different final outcomes.}

\begin{table}[t]
\begin{center}
\begin{tabular}{lllll}  \toprule
\textbf{Physics Systems} & $I$ & $J$ & $T$ & $r$ \\ \hline

1d Harmonic Oscillator & 200 & 20 & 1 & $10^4$  \\ \hline
2d Harmonic Oscillator & 50 & 20  & 2 & $10^4$  \\ \hline
Discrete Sine-Gordon & 100 & 50 & 2 & $10^4$\\ \hline
Calogero-Moser & 200 & 50 & 2 & $10^4$  \\ \hline
Toda lattice & 50 & 50 & 2 & 10  \\ \hline
FPUT & 1500 & 50 & 1 & 50  \\ \bottomrule
\end{tabular}
\end{center}
\caption{Parameters used for data generation. See Section~\ref{sec: data_generation} for a detailed description of the notations.}
\label{tab:data generation for systems}
\end{table}

\begin{table}[t]
\begin{center}
\begin{tabular}{lll}  \toprule
Number of Layers & 7  \\ \hline
Batch Size & 32  \\ \hline
Input/Output dimension of Each Layer & 1000  \\ \hline
Activation Function & \textit{tanh}  \\ \hline
Initial Learning Rate & 1e-5  \\ \hline
Scheduled Learning Rate for Fine-Tuning & 1e-9  \\ \hline
Initial Training Steps & 40000  \\ \hline
Fine-Tuning Steps & 20000  \\ \hline
Adam $\beta_1$ & 0.9  \\ \hline
Adam $\beta_2$ & 0.999  \\ \hline
Adam Weight Decay & 0  \\ \bottomrule
\end{tabular}
\end{center}
\caption{Hyperparameters for training the HNNs}
\label{tab:HNNs_setting}
\end{table}

\subsection{Results} \label{sec: systems_results}

We carefully examine the errors that can arise when learning a Hamiltonian system through HNN from system trajectories, with a particular focus on the 1D harmonic oscillator. We then further explore how these inaccuracies can impact the downstream neural deflation learning of the remaining conservation laws for the 2D harmonic oscillator and other Hamiltonian lattice systems.

\subsubsection{Harmonic Oscillators}

\textbf{1D harmonic oscillator}. We first examine the relative error of the HNN-approximated Hamiltonian $H_{\bxi^*}(\bq, \bp)$ compared to the ground truth Hamiltonian $H(\bq, \bp)$~\eqref{eq:hamiltonian_1d_osci}. Since the neural deflation method relies not on the value of the Hamiltonian itself but on the vector field $f(\bx) = \left(\nabla_{\bp}H, -\nabla_{\bq}H\right)$, the relative error is measured by
\begin{equation}
     \text{Relative error} = \frac{\left\|\left(\nabla_{\bp}H_{\bxi^*} , -\nabla_{\bq}H_{\bxi^*}\right)
     - \left(\nabla_{\bp}H , -\nabla_{\bq}H\right)\right\|}
     {\left\|\left(\nabla_{\bp}H , -\nabla_{\bq}H\right)\right\|}
     \label{eq:relative_error}
\end{equation}

\begin{figure}[h!]
    \centering
    \includegraphics[width=.6\textwidth]{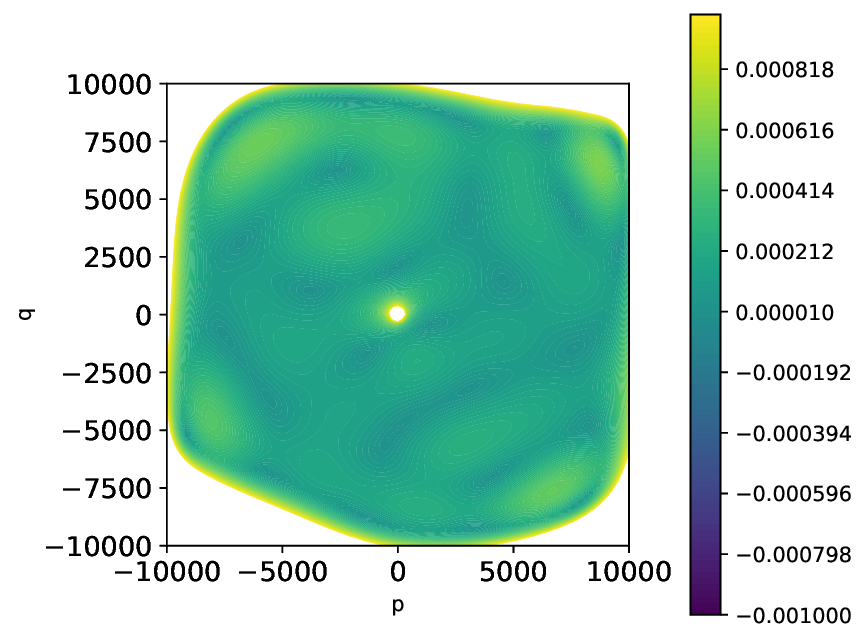}
    \caption{Relative error of HNNs, as defined in Eq.~\eqref{eq:relative_error}, evaluated over the domain \([-10^4, 10^4]\) for the 1D harmonic oscillator.} 
    \label{fig:1d_spring_hnn}
\end{figure}

Figure~\ref{fig:1d_spring_hnn} visualizes the relative error over the domain \([-10^4, 10^4]\), consistent with the domain used for sampling the training trajectories; see Table~\ref{tab:data generation for systems}. The HNN achieves high accuracy, with most relative errors within \(10^{-3}\), in approximating the true dynamics. This accuracy supports the application of the HNN approach to more complex systems, enhancing confidence in its effectiveness for the downstream neural deflation method.

\textbf{2D harmonic oscillator}.  We first train an HNN using data generated from the 2D oscillator system, similar to the 1D case. The learned Hamiltonian is then used as input for the neural deflation method, following Algorithm~\ref{alg:deflation}. As part of an ablation study, we also explore the effects of varying deflation strength, using \(\alpha = 1.0\) and \(\alpha = 0.5\) in Eq.~\eqref{eq:lK}.

\begin{figure}[h!]
    \centering
    \begin{overpic}[width=0.45\textwidth]{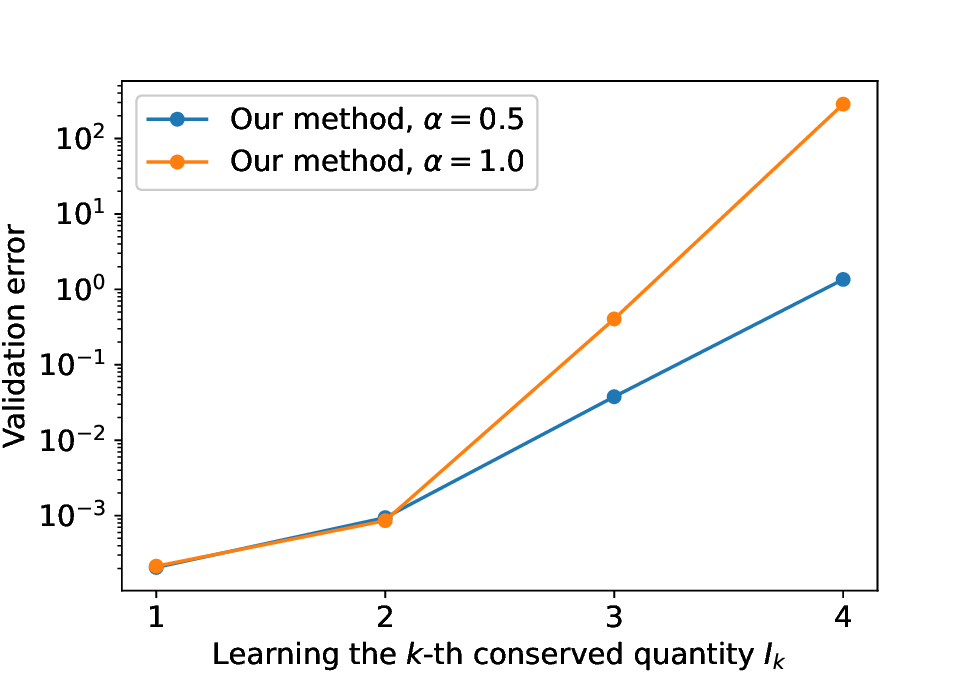}
    \put(50,-4){$(a)$}
     \end{overpic}
     \begin{overpic}[width=0.45\textwidth]{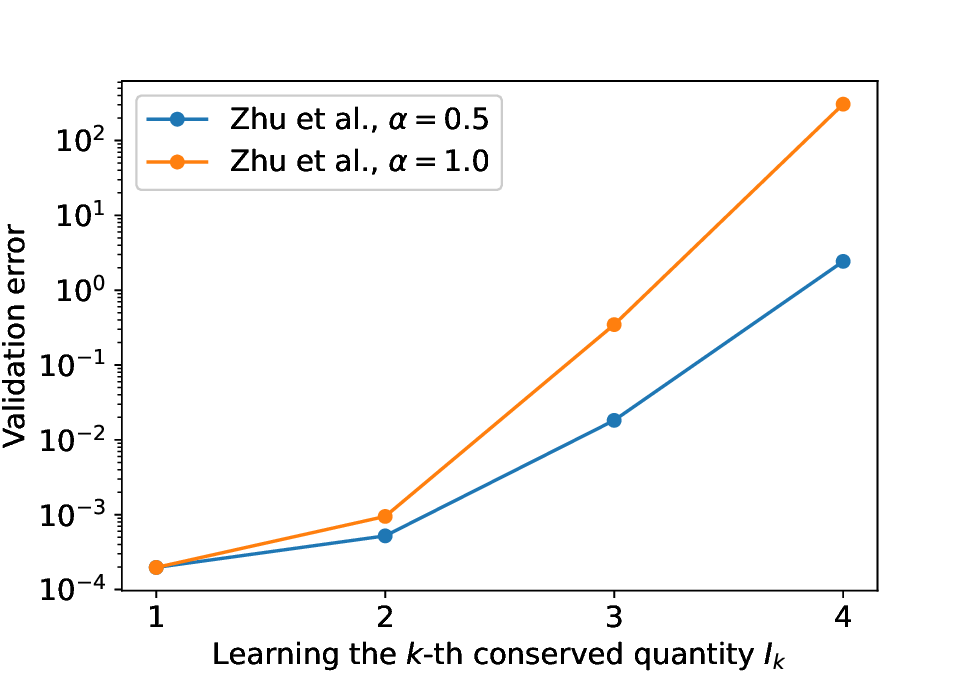}
    \put(50,-4){$(b)$}
     \end{overpic}
    \vspace{0.2cm}
    \caption{2D harmonic oscillator. This figure shows the validation losses \(\{\mathcal{L}_k(\bm{\theta}^{*}_k ; \mathcal{V})\}^{4}_{k=1}\) for the learned conserved quantities \(\{I_k(\cdot; \bm{\theta}^{*}_k) \}^{4}_{k=1}\) under different deflation strengths (\(\alpha=1.0\) and \(\alpha=0.5\)). Results are compared for: (a) our model, which learns solely from system trajectories, and (b) the original neural deflation method~\cite{neural_deflation}, which uses explicit knowledge of the ground truth differential equation. Both methods exhibit a significant increase in loss at \(k=3\), indicating successful identification of the system's integrability, as per Algorithm~\ref{alg:deflation}.}
    \label{fig:2d_spring}
\end{figure}

Figure~\ref{fig:2d_spring} compares (a) our model, which learns exclusively from system trajectories, with (b) the original neural deflation method~\cite{neural_deflation}, which uses additional explicit knowledge of the ground truth differential equation. Both approaches exhibit a sharp increase in the validation loss $\mathcal{L}_k(\mathbf{\btheta}^{*}_k ; \mathcal{V})$, by several orders of magnitude, at \(k=3\),
depending also on the value of the deflation exponent
$\alpha$. This result suggests that both methods have correctly identified the system's integrability (as indicated in the final step of Algorithm~\ref{alg:deflation}) and successfully discovered a maximal set of (two) independent conservation laws that are in involution. Given that a larger deflation strength of $\alpha=1.0$ results in a more pronounced jump in loss, we will use this value of $\alpha$ for the remaining examples.

\subsubsection{Discrete Sine-Gordon System}
\begin{figure}[h!]
    \centering
    \begin{overpic}[width=0.45\textwidth]{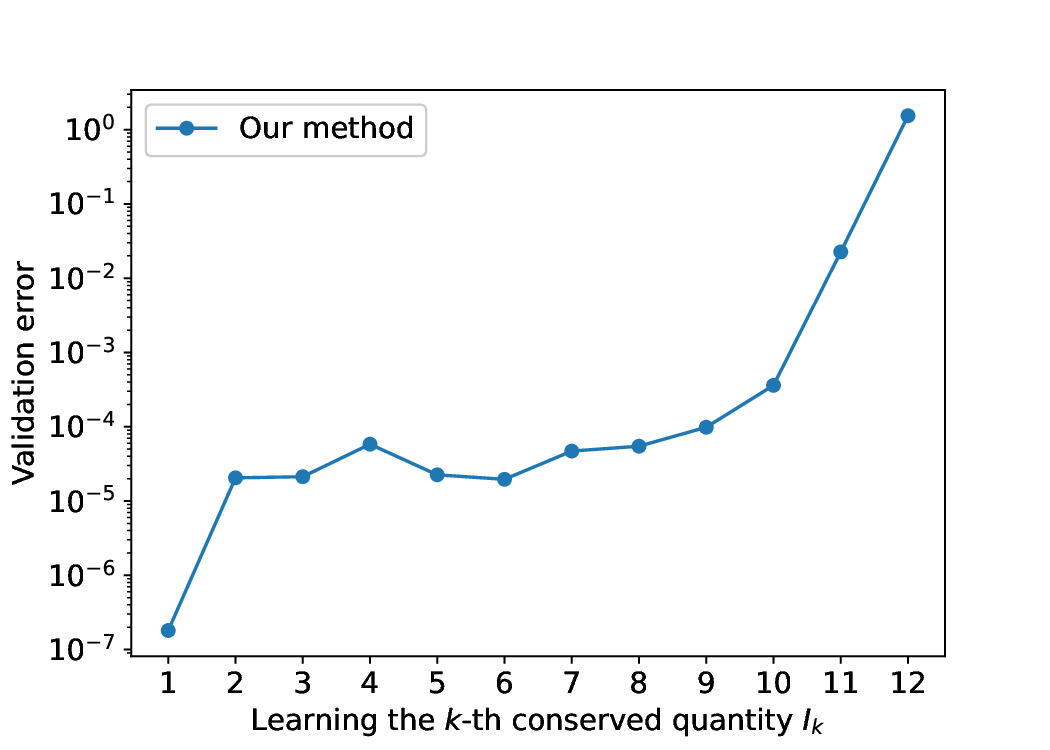}
    \put(50,-4){$(a)$}
     \end{overpic}
     \begin{overpic}[width=0.45\textwidth]{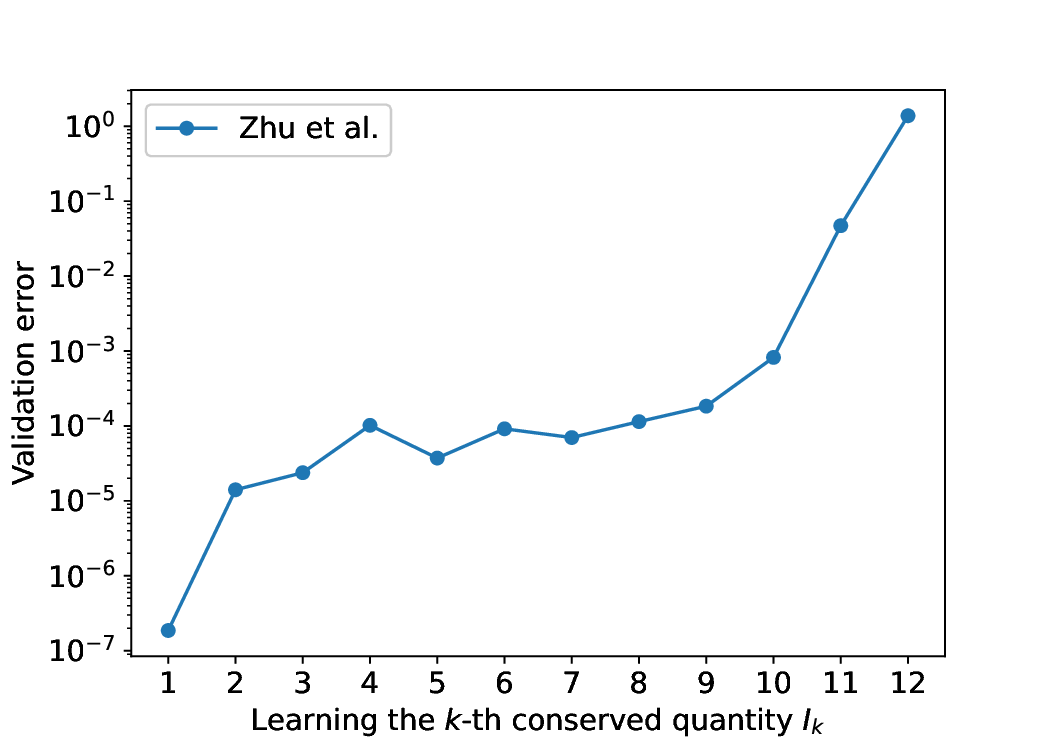}
    \put(50,-4){$(b)$}
     \end{overpic}
    \vspace{0.2cm}
    \caption{Discrete sine-Gordon system with the number of lattice sites/degrees of freedom $N=d=6$. This figure shows the validation losses \(\{\mathcal{L}_k(\bm{\theta}^{*}_k ; \mathcal{V})\}^{2d}_{k=1}\) for the learned conserved quantities \(\{I_k(\cdot; \bm{\theta}^{*}_k) \}^{2d}_{k=1}\). Results are compared for: (a) our model, which learns solely from system trajectories, and (b) the original neural deflation method~\cite{neural_deflation}, which uses explicit knowledge of the ground truth differential equation. Both methods exhibit a significant increase in loss at \(k=2\), consistent with the fact that the underlying system has only \textit{one} independent conservation law; see Algorithm~\ref{alg:deflation}.}
    \label{fig:sine_gordon}
\end{figure}

Figure~\ref{fig:sine_gordon} offers a side-by-side comparison of our model with the original neural deflation method~\cite{neural_deflation} applied to the non-integrable discrete sine-Gordon system, with the number of lattice sites (and thus the degrees of freedom) set to \(N=6=d\). The results are similar, despite our model relying solely on simulated system trajectories. Both methods exhibit a significant increase in validation loss at \(k=2\), by about 2 orders
of magnitude, consistent with the fact that the underlying system has only one independent conservation law (refer to the last line of Algorithm~\ref{alg:deflation}).

\subsubsection{Calogero-Moser System} 

\begin{figure}[h!]
    \centering
    \begin{overpic}[width=0.45\textwidth]{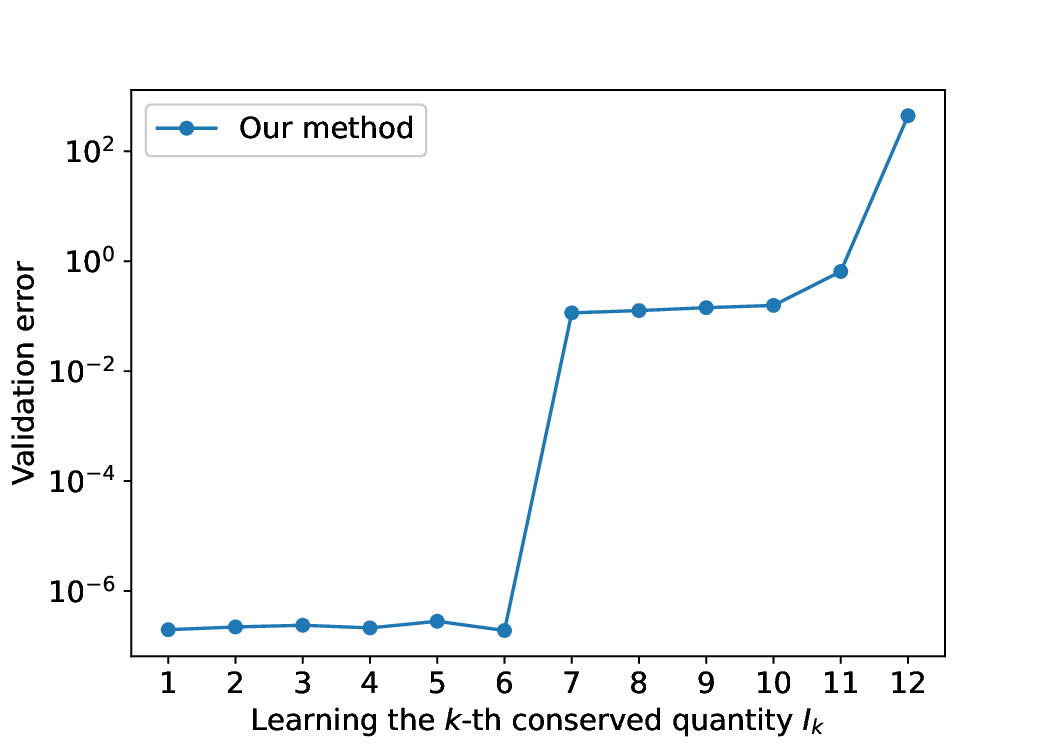}
    \put(50,-4){$(a)$}
     \end{overpic}
     \begin{overpic}[width=0.45\textwidth]{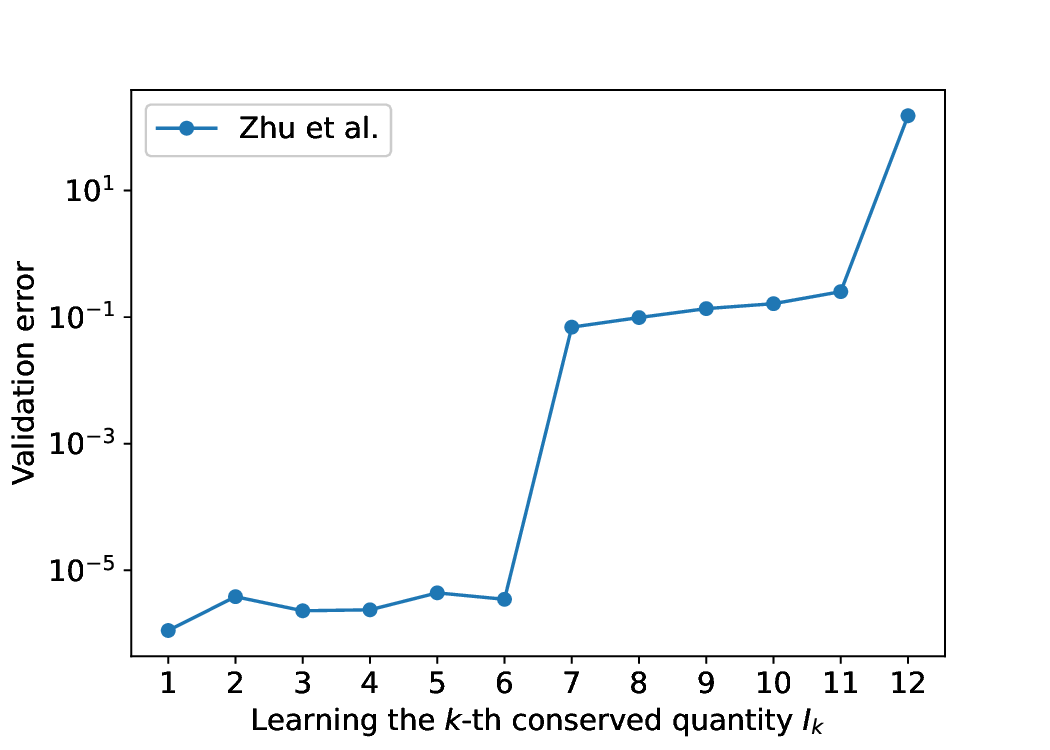}
    \put(50,-4){$(b)$}
     \end{overpic}
    \vspace{0.2cm}
    \caption{Calogero-Moser system with the number of lattice sites/degrees of freedom $N=d=6$. This figure shows the validation losses \(\{\mathcal{L}_k(\bm{\theta}^{*}_k ; \mathcal{V})\}^{2d}_{k=1}\) for the learned conserved quantities \(\{I_k(\cdot; \bm{\theta}^{*}_k) \}^{2d}_{k=1}\). Results are compared for: (a) our model, which learns solely from system trajectories, and (b) the original neural deflation method~\cite{neural_deflation}, which uses explicit knowledge of the ground truth differential equation. Both methods \textbf{appear} to exhibit a significant increase in loss at \(k=7\), consistent with the fact that the underlying system is fully-integrable.}
    \label{fig:Calogero}
\end{figure}

Figure~\ref{fig:Calogero} presents the results when the methods are applied to the fully integrable Calogero-Moser system with \(N=6\) lattice sites. At first glance, our method [panel (a)] achieves comparable, if not superior, performance compared to the original neural deflation method [panel (b)]. Both methods exhibit a significant jump in test loss precisely at \(k=7\), indicating the successful learning of six independent, Poisson-commuting conservation laws in this integrable system with six degrees of freedom. Notably, the jump in test loss is even more pronounced with our method compared to the original neural deflation approach.

\begin{figure}[h!]
  \centering
  \includegraphics[width=0.55\linewidth]{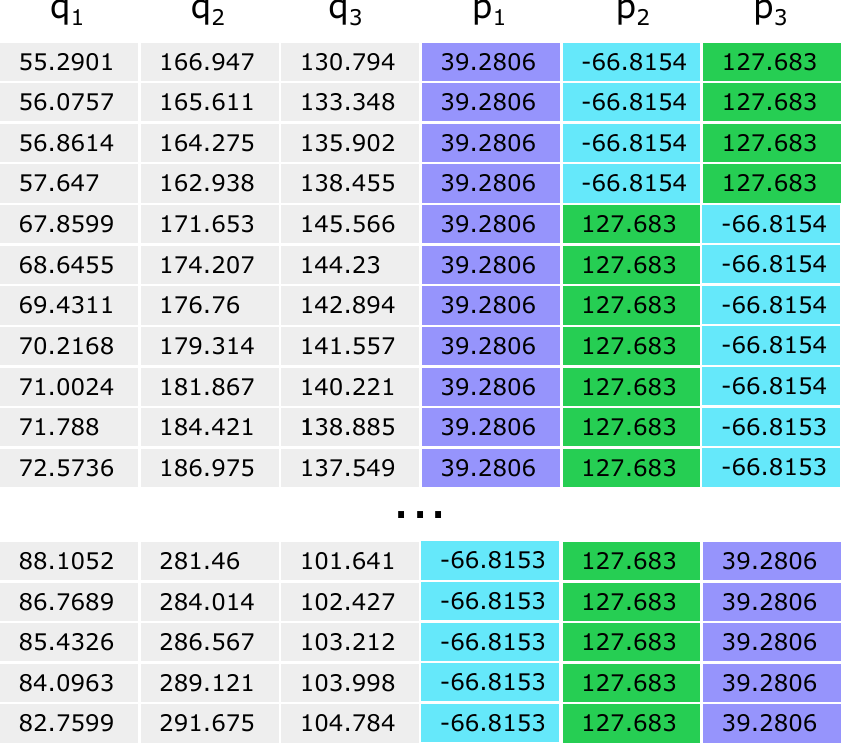}
   \caption{A simulated trajectory from the Calogero-Moser system with $N = d = 3$ degrees of freedom. Each column presents the time evolution of the dynamic variables $q_i$ and $p_i$. The particles move at constant speeds, exchanging momentum almost instantaneously during collisions (when $q_i \approx q_j$), due to short-range interactions governed by an inverse cubic force. This momentum exchange is indicated by the sudden change of colors in the columns for $p_i$. To highlight the moments of momentum exchange, some time data has been omitted from the illustration.}
   \label{fig:calogero_traj}
\end{figure}

However, a closer examination of the Hamiltonian learned by the HNN reveals some unexpected behavior. Due to the short-range interactions in the Calogero-Moser system, which are governed by an inverse cubic force, most particles maintain constant speeds and only exchange momentum almost instantaneously during brief collisions (i.e., when \(q_i \approx q_j\)). This phenomenon is illustrated in Figure~\ref{fig:calogero_traj}, {where the (rare) momentum exchange is indicated by the sudden change of colors in the columns for $p_i$.} As a result, the time-derivative data \(\dot{\mathcal{D}}\), obtained using finite differences, makes the system behave similarly to one with zero potential, with rare ``outliers" occurring during particle collisions. Consequently, the HNN tends to approximate the system Hamiltonian as having zero potential energy.

A potential solution to this issue is to use neural ODEs~\cite{Alpher06} instead of HNNs to learn the underlying Hamiltonian, thus avoiding the inaccuracies introduced by finite difference estimation of time derivatives. This approach will be explored in future work.

\subsubsection{Toda system}
\label{sec:toda_experiments}

\begin{figure}[h!]
    \centering
    \begin{overpic}[width=0.45\textwidth]{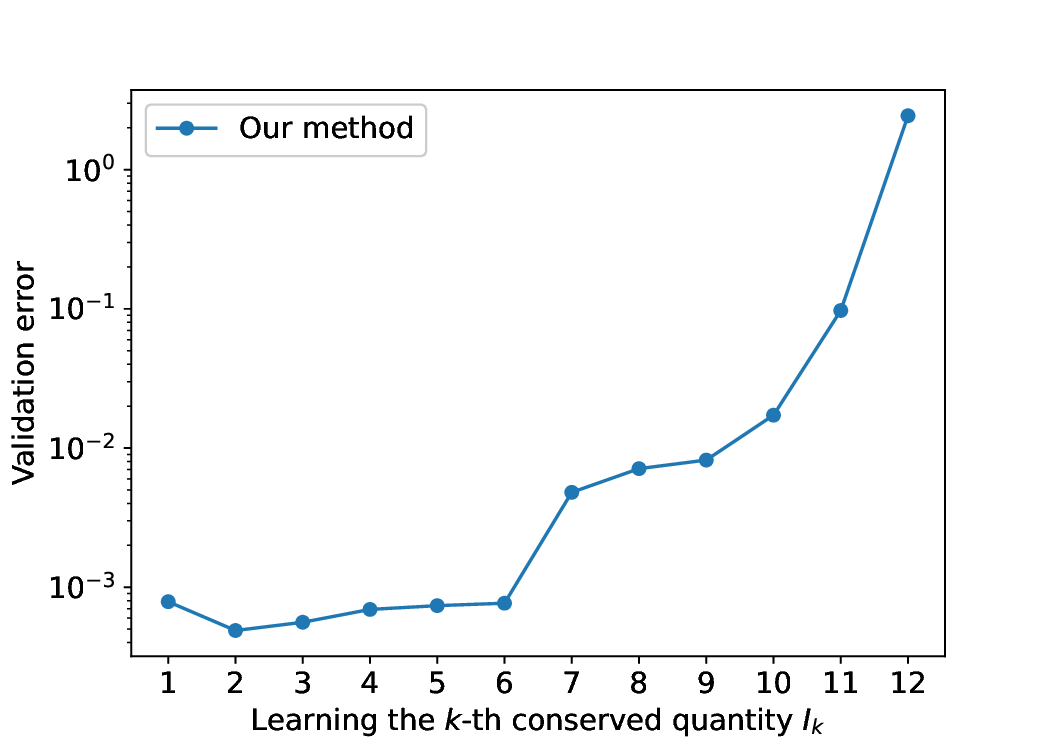}
    \put(50,-4){$(a)$}
     \end{overpic}
     \begin{overpic}[width=0.45\textwidth]{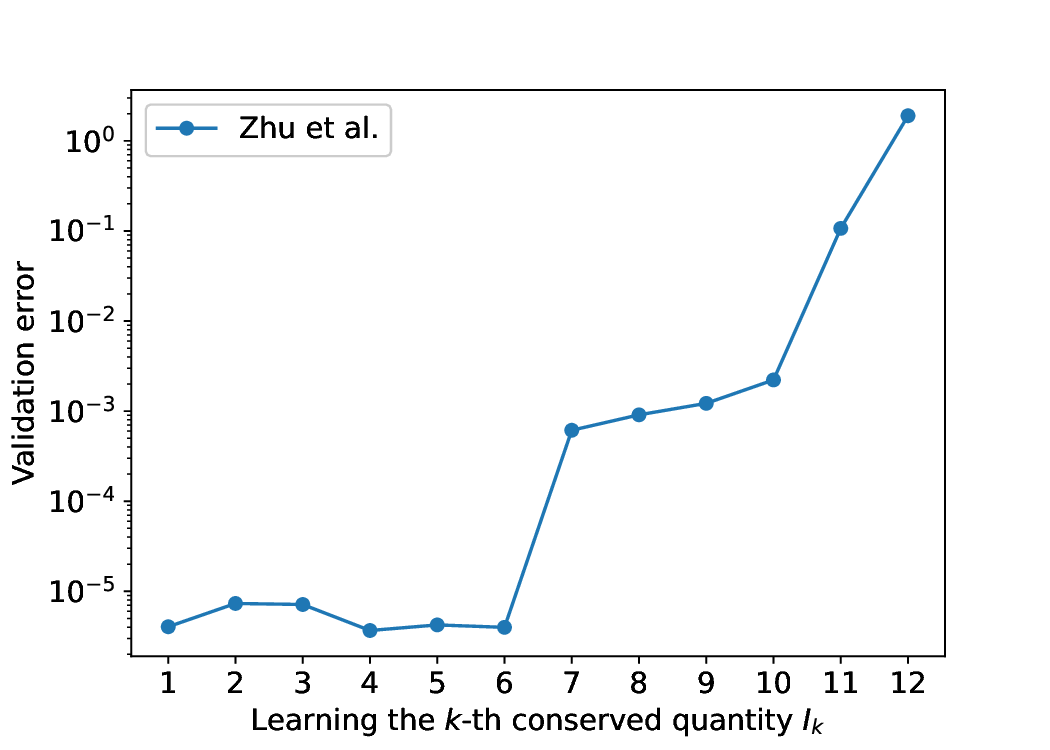}
    \put(50,-4){$(b)$}
     \end{overpic}
    \vspace{0.2cm}
    \caption{Toda system with the number of lattice sites/degrees of freedom $N=d=6$. This figure shows the validation losses \(\{\mathcal{L}_k(\bm{\theta}^{*}_k ; \mathcal{V})\}^{2d}_{k=1}\) for the learned conserved quantities \(\{I_k(\cdot; \bm{\theta}^{*}_k) \}^{2d}_{k=1}\). Results are compared for: (a) our model, which learns solely from system trajectories, and (b) the original neural deflation method~\cite{neural_deflation}, which uses explicit knowledge of the ground truth differential equation. Both methods show a notable increase in validation loss at \(k=7\), indicating the successful identification of \(d_0=7-1=6\) independent conservation laws in involution, thereby confirming the integrability of the Toda system. The jump in validation loss is more pronounced in panel (b), where the original neural deflation method incorporates explicit information about the underlying Hamiltonian~\cite{neural_deflation}. The reduced sharpness of the loss jump in our model likely stems from the exponential term in the Toda system's potential~\ref{eq:toda_lattice}, which complicates accurate learning of the Hamiltonian.}
    \label{fig:toda}
\end{figure}

We now examine the integrable Toda system with degrees of freedom \(d = N = 6\), as illustrated in Figure~\ref{fig:toda}. Similar to the original neural deflation method shown in panel (b), our method in panel (a) exhibits a significant increase in validation loss at \(k=7\), indicating the successful identification of \(d_0 = 7 - 1 = 6\) independent conservation laws in involution, thereby confirming the system's integrability. However, the jump in validation loss is more pronounced in the original neural deflation method~\cite{neural_deflation}, which explicitly incorporates the Hamiltonian into the algorithm. This discrepancy likely arises from the exponential term in the Toda lattice potential~\eqref{eq:toda_lattice}, which complicates the high-accuracy approximation of the Hamiltonian by HNN over a large domain. Nevertheless, the similarity in loss behavior between our data-driven method, which learns conservation laws from system trajectories, and the original neural deflation method, which assumes explicit knowledge of the underlying ODE system, highlights the potential of our approach for learning conservation laws and assessing system integrability from observed trajectories.

\subsubsection{FPUT System}
\label{sec:fput_experiments}
\begin{figure}[h!]
    \centering
    \begin{overpic}[width=0.45\textwidth]{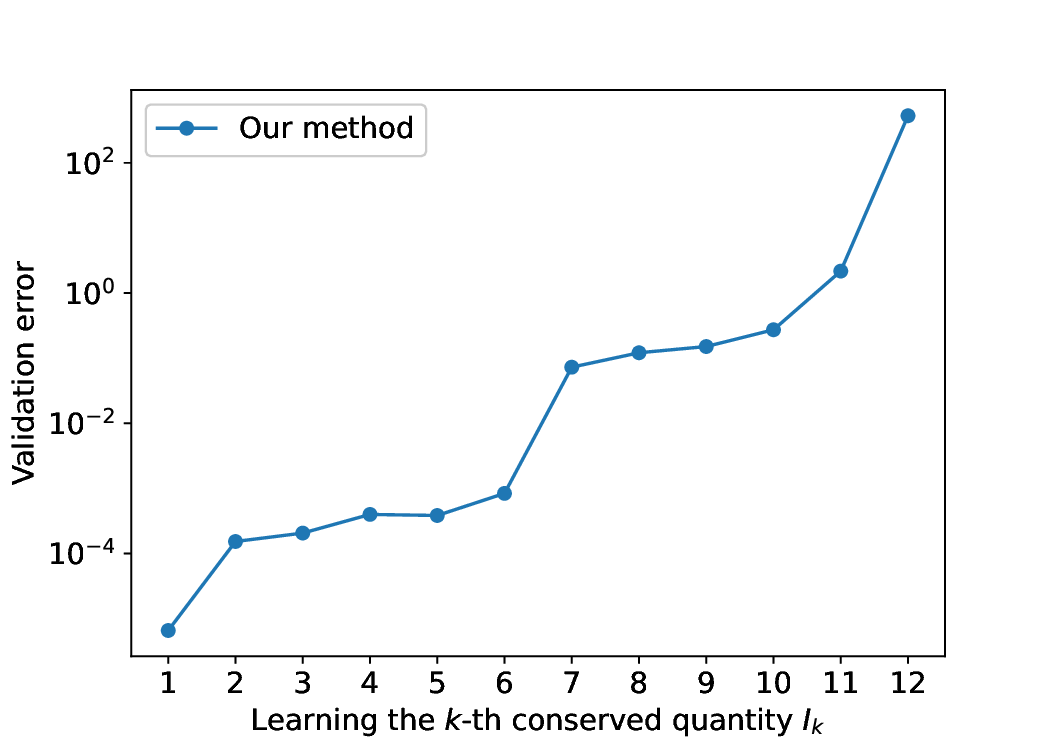}
    \put(50,-4){$(a)$}
     \end{overpic}
     \begin{overpic}[width=0.45\textwidth]{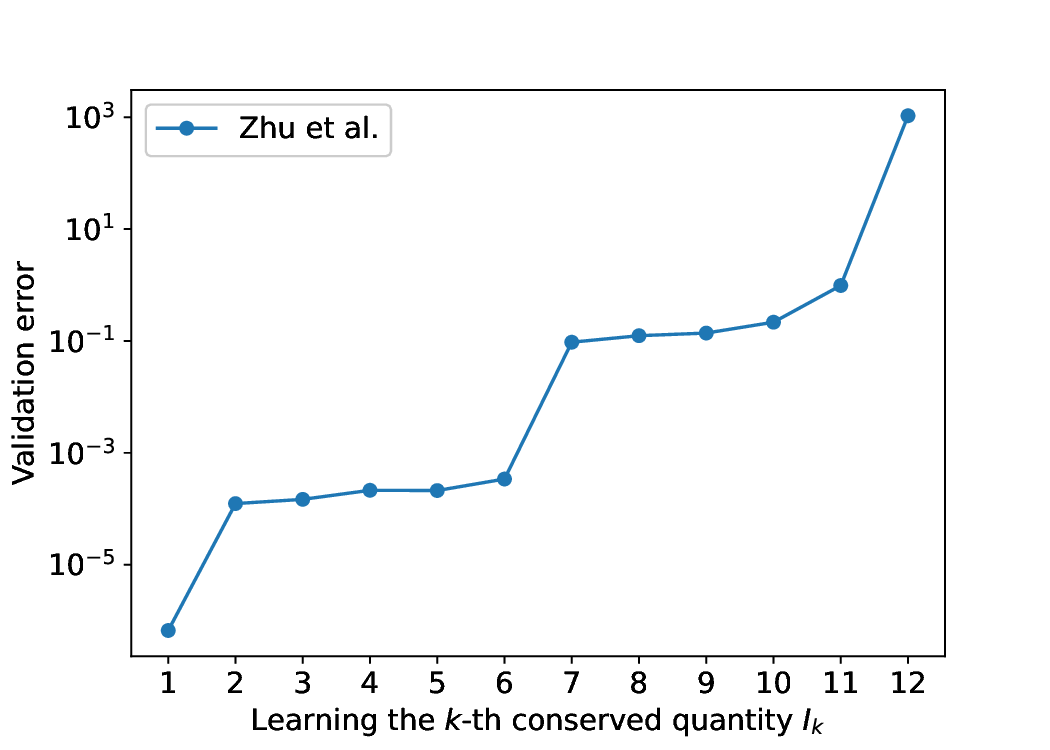}
    \put(50,-4){$(b)$}
     \end{overpic}
    \vspace{0.2cm}
    \caption{FPUT system with the number of lattice sites/degrees of freedom $N=d=6$. This figure shows the validation losses \(\{\mathcal{L}_k(\bm{\theta}^{*}_k ; \mathcal{V})\}^{2d}_{k=1}\) for the learned conserved quantities \(\{I_k(\cdot; \bm{\theta}^{*}_k) \}^{2d}_{k=1}\). Results are compared for: (a) our model, which learns solely from system trajectories, and (b) the original neural deflation method~\cite{neural_deflation}, which uses explicit knowledge of the ground truth differential equation. Both methods exhibit a significant increase in loss at \(k=2\), indicating the non-integrability nature of the FPUT system. See Section~\ref{sec:fput_experiments} for a detailed discussion of this result.}
    \label{fig:FPUT}
\end{figure}

Finally, we turn to the non-integrable FPUT system with the number of degrees of freedom remaining \(d = N = 6\), as illustrated in Figure~\ref{fig:FPUT}. Similar to the original neural deflation method in panel (b), our approach in panel (a) shows a clear increase in validation loss after identifying the first conservation law, indicating the non-integrable nature of the system. Unlike  the Toda system, where the jump in validation loss differed significantly between methods (see Section~\ref{sec:toda_experiments}), the increase here is of comparable magnitude. This is likely because the polynomial potential in the FPUT system is much easier for the HNN to approximate accurately. However, consistent with \cite{neural_deflation}, our method only detects one conservation law, even though the FPUT system has two independent conservation laws in involution (momentum and Hamiltonian). It remains an open question whether a modified neural deflation method could correctly identify both. Despite this, the comparable validation loss patterns between our method and the original neural deflation approach suggest that our method shows strong potential for effectively learning conservation laws and evaluating system integrability using only trajectory data.

\section{Conclusions and Future Works}
\label{sec:conclusion}
In this work, we introduced a data-driven method for discovering independent conservation laws in Hamiltonian systems using neural deflation, relying solely on system trajectory data. Our method successfully identified the (almost) exact number of conservation laws in both integrable and non-integrable systems, demonstrating performance comparable to prior state-of-the-art approaches that require explicit knowledge of the underlying ODEs. These results underscore the potential of our approach for learning conservation laws from directly observed data that may, in principle, arise even from a 
physical experiment.

%Future work will focus on 
A promising focal point for future work consists of improving the accuracy of data-driven Hamiltonian learning for systems with complex dynamics, such as the Toda and Calogero-Moser systems. In particular, in the Calogero-Moser system, short-range interactions cause particles to maintain constant speeds, with momentum exchanges occurring only during brief collisions (in a way
reminiscent of hard sphere models), making finite difference estimates of time derivatives unreliable. To address this, a promising avenue involves
exploring neural ODEs in place of HNNs, which can avoid errors from finite differences and remove the need for short time intervals in trajectory data. 
On the other hand, it is also an open problem whether the momentum conservation of FPUT models can be identified in the context of the methods proposed herein.
Additionally, we plan to use symbolic regression to connect the learned conservation laws to known physical quantities or those identified through integrability techniques, enhancing interpretability.
There exist also interesting perspectives regarding connecting Liouville with
other types of integrability (e.g., Frobenius)~\cite{liu2022machine}, as well
as in connection to identifying the symplectic transformations leading
to the action-angle variables associated with the conserved quantities~\cite{bondesan2019learning}.
These directions are currently under investigation and will be detailed in future work.

\section*{Acknowledgements}
This material is based upon work supported by the U.S. National Science Foundation under awards DMS-2052525, DMS-2140982, DMS-2244976 (WZ), PHY-2110030, PHY-2408988, and DMS-2204702 (PGK), as well as DMS-2220211 (HZ). Additional support was provided by the Simons Foundation Collaboration Grants for Mathematicians 706383 (HZ).

\newpage 

\newpage

\bibliographystyle{unsrt}
\bibliography{main}

\end{document}